# Expressing an NP-Complete Problem as the Solvability of a Polynomial Equation

Deepak Ponvel Chermakani, IEEE Member
deepakc@pmail.ntu.edu.sg  deepakc@usa.com  deepakc@myfastmail.com  deepak.chermakani@ustri.com

*Abstract*:  We demonstrate a polynomial approach to express the decision version of the directed Hamiltonian Cycle Problem (HCP), which is NP-Complete, as the Solvability of a Polynomial Equation with a constant number of variables, within a bounded real space. We first introduce four new Theorems for a set of periodic Functions with irrational periods, based on which we then use a trigonometric substitution, to show how the HCP can be expressed as the Solvability of a single polynomial Equation with a constant number of variables. The feasible solution of each of these variables is bounded within two real numbers. We point out what future work is necessary to prove that P=NP.

## 1.   Introduction

Given a graph $G\ (V,\ E)$, where $V$ is the set of $N$ vertices, and $E$ is the set of edge-costs that are either 1 or 0, the decision version of the HCP is to find out whether there exists a round trip tour to visit each vertex exactly once, such that the total cost of the tour equals $N$. If the edge-cost in traveling from one edge to another, is the same as the reverse cost, then the HCP is said to be undirected, while if it is either same or different than the reverse cost, then the HCP is said to be directed. Both the directed, as well as the undirected forms of the decision version of the HCP, are known to be NP-Complete.

   In the rest of this paper, we shall refer to "decision version of the directed HCP", simply as the "HCP". Our paper describes a new polynomial Approach for expressing the HCP as the solvability of an Equation. Our Approach is discussed in Section-3 and is based on new Theorems, which we will introduce in the next section (Section-2) of this paper.

## 2.   Theorems on a Set of Periodic Functions with Irrational Periods

### 2.1   Definition of the Periodic Functions

Let us define $n$ periodic Functions to have the following three properties.

**Property-1**: Period of Function $i$, $T_i = (1 / (1 + iR))$, where $R$ is irrational non-zero real, and $i$ represents integers 1 to $n$

**Property-2**: All the $n$ Functions are binary functions and have a binary *high* value for the same duration $x$, where $x$ is non-zero, real, positive, and lesser than the smallest period, which is $1 / (1 + nR)$.

**Property-3**: Function $i$ is allowed to start to become high from time $t=0$, with real delay $D_i$, where $0 \leq D_i < T_i$

   For better clarity, we attempt to show these three properties of the $n$ Functions, in Figure-1 below. Note that the variable $t$ along the t-Axis, varies from 0 to positive infinity.

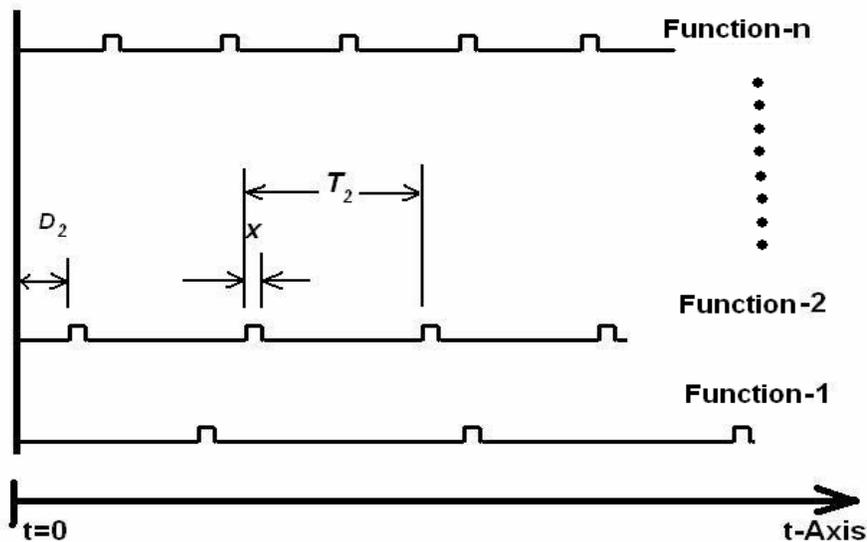

**Fig 1.    Definition of the Periodic Functions**

## 2.2 Definition of $E(t)_i$

We define $E(t)_i$ as the closest positive displacement from $t$ when Function $i$ again starts to become *high*. For the purpose of better clarity, we show this definition in Figure 2 below.

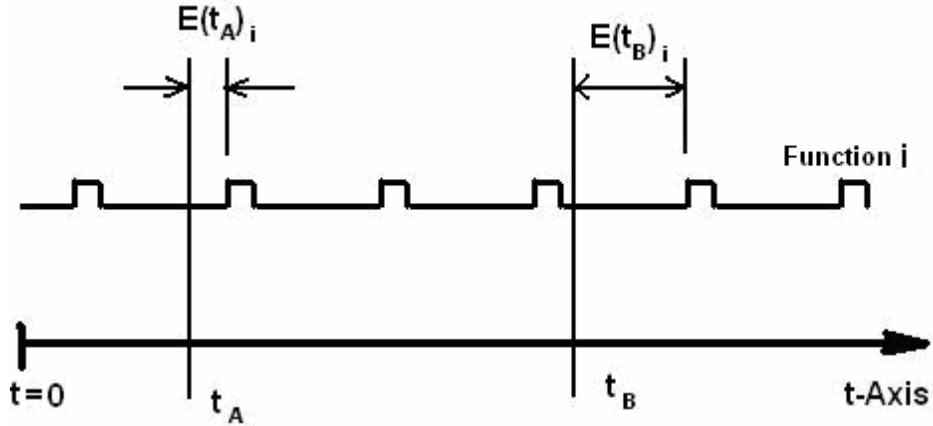

**Fig 2.** Definition of $E(t)_i$, for Function $i$

## 2.3 Theorem-1:

If we consider any two Functions $i$ and $j$ from the $n$ Functions defined in Section 2.1, then every real value of $t$ in $[0, \infty[$ has a unique set $\{E(t)_i, E(t)_j\}$. In other words, $\{E(t_A)_i, E(t_A)_j\} \neq \{E(t_B)_i, E(t_B)_j\}$, if $t_A \neq t_B$.

**Proof**: If $\{E(t_A)_i, E(t_A)_j\} = \{E(t_B)_i, E(t_B)_j\}$, for say $t_A < t_B$, then it would mean that it is possible to identify another point $t_C$ where $\{E(t_B)_i, E(t_B)_j\} = \{E(t_C)_i, E(t_C)_j\}$, and where $t_C - t_B = t_B - t_A$. It would also mean that an infinite number of similar points like $t_C$ can be periodically found along the t-Axis. This, in turn, would imply that $N_1 T_i = N_2 T_j$, for some natural numbers $N_1$ and $N_2$, where $N_1$ and $N_2 \neq 0$. But we know that $(T_i / T_j) = (1 + jR) / (1 + iR)$, which is irrational and therefore cannot be expressed as $(N_2 / N_1)$. **Hence Proved Theorem-1**.

## 2.4 Theorem-2:

If we consider all the $n$ Functions defined in Section 2.1, then every real value of $t$ in $[0, \infty[$ has a unique
set $\{E(t)_1, E(t)_2, E(t)_3, \ldots, E(t)_{n-1}, E(t)_n\}$

**Proof**: This is straightforward from Theorem-1, and it is easy to see that all the periods of the $n$ Functions are irrational with respect to each other. **Hence Proved Theorem-2**.

## 2.5 Theorem-3:

If we consider any two Functions $i$ and $j$ from the $n$ Functions defined in Section 2.1, then the following two statements are True (also see Figure 3):

i) There exists an infinite number of Intervals along t-Axis, where these two functions are simultaneously *high* for a duration of atleast $(x - \mu)$, where $\mu$ is positive non-zero real and $(\mu/x) \rightarrow 0$.

ii) The starting points of such Intervals (e.g. $t_A$, $t_B$ in Figure 3) do not have a constant period, but their occurrence along the t-Axis can be expressed by some relationship involving $T_i$, $T_j$, $D_i$, $D_j$, $\mu$, and $x$.

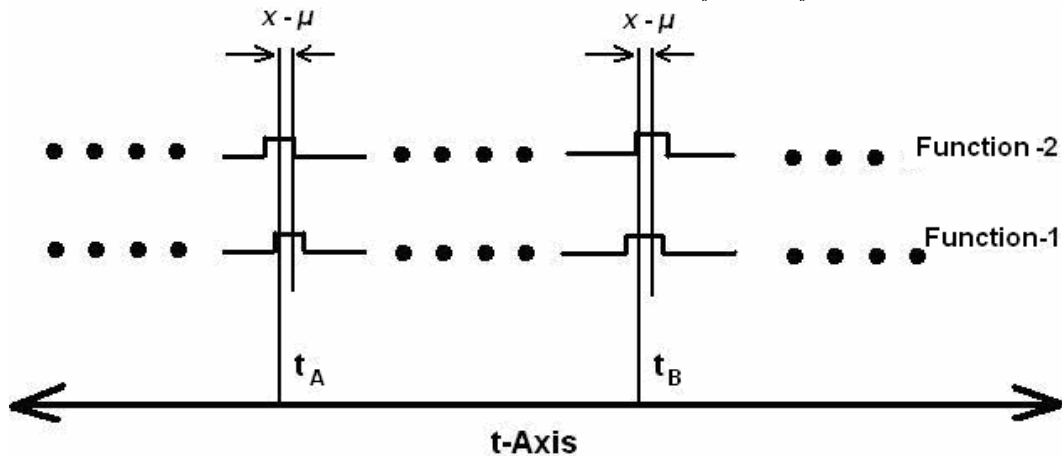

**Fig 3.** There are infinite Intervals along t-Axis, when the 2 Functions are simultaneously high for duration of atleast $x - \mu$

**Proof:** We will first prove this by assuming zero-delays for Functions *i* and *j*. We will then incorporate delays, and show that the proof holds. We define a 'Token', as a movable point used to indicate the current location of a process along the t-Axis.

<u>Lemma-3.1</u>: If $D_i = D_j = 0$, there exists some real t > 0, where the start of the *highs* of Functions *i* and *j*, come closer to each other upto a small positive distance $\mu$, where $(\mu/x) \rightarrow 0$.
<u>Proof</u>: Consider two tokens, Tokens *A* and *B*, to be placed at t=0. If $T_i > T_j$, then let $T_A = T_i$ and $T_B = T_j$, else let $T_A = T_j$ and $T_B = T_i$. Now follow the procedure below (note that INT (argument) returns Integer portion of that argument):
Step 1: Set *Position_of_Token_A* = $T_A$
Step 2: Set *Position_of_Token_B* = *Position_of_Token_B* + $T_B$
Step 3: Go back to Step 2, if Token *B* has not yet overtaken Token *A*
Step 4: Set *Difference_in_token_positions* = ( *Position_of_Token_B* - *Position_of_Token_A* )
Step 5: Set *Integral_Multiple* = 1 + INT ( (*Position_of_Token_A*) / (*Difference_in_token_positions*) )
Step 6: Set *Position_of_Token_A* = ( 2 x *Position_of_Token_A* ) x ( *Integral_Multiple* )
Step 7: Set *Position_of_Token_B* = *Position_of_Token_B* x ( *Integral_Multiple* )
Step 8: Go back to Step 4
     It will be observed that the variable *Difference_in_token_positions* will continue to decrease to a very small value (though it will never become zero), with each pass through Step 4 of the procedure. There are two reasons why this happen. First, the remainder of division is less than the divisor during the division process (see Step 5), and in the case of our Procedure, the k[th] time that Token *B* overtakes Token *A*, then the divisor for the (k+1)[th] overtaking is being obtained from the remainder (i.e. difference in Token positions) of the k[th] overtaking. Second, a zero remainder is prevented from occurring during the division process, because it is impossible for ( (*Position_of_Token_A*) / (*Difference_in_token_positions*) ) to be an integer. The key factor which enables this happen is the fact that the ratio of periods of Functions *i* and *j*, is irrational.
     Next, as it is obvious that the positions of both Tokens *A* and *B*, represent the start of *highs* of Functions *i* or *j* (because both Tokens start from t=0, and because both Tokens are iteratively shifted forward by integral multiples of $T_i$ and $T_j$), we can conclude that at some real value of t > 0, the start of *highs* of Functions *i* and *j*, come closer to each other, upto a small positive distance $\mu$, where $(\mu/x) \rightarrow 0$. It is also important to note that the start of *highs* will never touch each other (i.e. $\mu \neq 0$), because if they touch, then it would mean that $N_1 T_i = N_2 T_j$, for some natural numbers $N_1$ and $N_2$, where $N_1$ and $N_2 \neq 0$, which is false as we proved earlier in Theorem-1. <u>Hence Proved Lemma-3.1</u>

<u>Lemma-3.2</u>: If $D_i = D_j = 0$, then there exist an infinite number of Intervals along t-Axis, where Functions *i* and *j*, are simultaneously *high* for a duration of atleast $(x - \mu)$, where $\mu$ is positive non-zero real and $(\mu/x) \rightarrow 0$. Further, the starting points of such Intervals (e.g. $t_A$, $t_B$ in Figure 3) do not have a constant period, but their occurrence along the t-Axis can be expressed by some relationship involving *t*, $T_i$, $T_j$, $\mu$, and *x*.
<u>Proof</u>: We just proved in Lemma-3.1 that the start of the *highs* of Functions *i* and *j*, do come closer to each other, upto a small positive distance $\mu$, where $(\mu/x) \rightarrow 0$. Let $L_i$ denote the distance from t=0 to the start of that *high* of Function *i*, and let $L_j$ denote the distance from t=0 to the start of that *high* of Function *j*, so that abs $(L_i - L_j) = \mu$. Now starting from t=0, let us place infinite Tokens of type A along the t-Axis so that two neighboring tokens of type A are separated by distance $L_i$. Similarly, starting from t=0, let us place infinite Tokens of type B along the t-Axis so that two neighboring tokens of type B are separated by distance $L_j$. Now imagine a single Token C, starting from t=0, and iteratively taking leaps of $L_i$ every unit of time. It is obvious that Token C will observe a discrete relative-speed (i.e. discrete steps of $\mu$ per unit time) of an infinite series of Tokens of type B. This can also be imagined as if stationary Token C observes a moving Train having infinite carriages, and where the start of each carriages is represented by a Token of type B, moving in the opposite direction (if $L_i > L_j$) or moving in the same direction (if $L_i < L_j$), with a discrete relative-speed (i.e. discrete steps of $\mu$ per unit time).
     As $(\mu/x) \rightarrow 0$ (which means $\mu$ is very much greater than *x*), it can be concluded that there exists an infinite number of Intervals along t-Axis, where Functions *i* and *j*, are simultaneously *high* for a duration of atleast $(x - \mu)$, where $\mu$ is positive non-zero real and $(\mu/x) \rightarrow 0$. It is important to note that the starting points of such Intervals cannot have a constant period. It is easy to see that if they do have a constant period, then it would mean that $N_1 T_i = N_2 T_j$, for some natural numbers $N_1$ and $N_2$, where $N_1$ and $N_2 \neq 0$, which is false as we earlier proved. Instead, the occurrence of such Intervals along the t-Axis is governed by a very sophisticated relationship among *t*, $T_i$, $T_j$, $\mu$, and *x*. It is beyond our scope to go into details of that relationship. The existence of such a relationship is verified since the occurrence of the starting points of such Intervals is more frequent, if the values of $\mu$ and *x* are increased. <u>Hence Proved Lemma-3.2</u>

     Finally, introducing delays $D_i$ and $D_j$, simply shifts the initial position of the infinite 'Train' of Tokens of type B, either forward or backward by *abs* $(D_i - D_j)$. Token C continues to observe the 'Train' moving with the same discrete relative-speed. So the Lemma-3.1 and Lemma-3.2 continue to remain true. **Hence Proved Theorem 3.**

## 2.6 Theorem-4:

If we consider all *n* Functions defined in Section 2.1, there are an infinite number of Intervals along the t-Axis, when all *n* Functions are simultaneously *high* for duration greater than $(x - \mu)$, where $(\mu/x) \to 0$, and $\mu \to 0$. Please see figure 4 below.

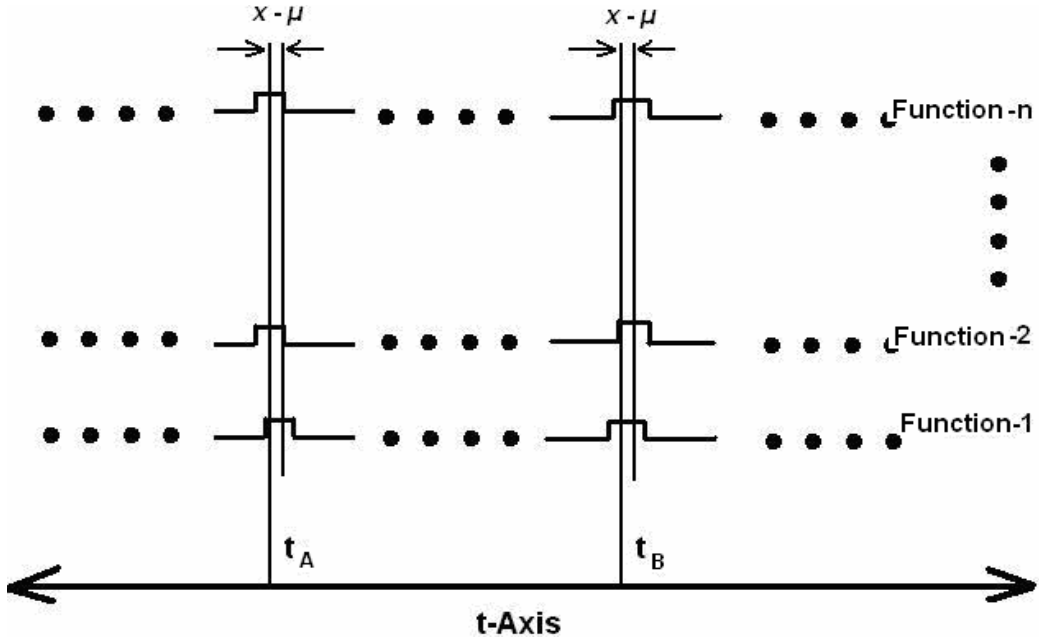

**Fig 4.** There are an infinite number of Intervals, when all *n* Functions are simultaneously *high* for duration greater than $x - \mu$

**Proof:** We shall give two angles of argument, to support the validity of this Theorem.

Angle-1: As discussed under the proof for Theorem-3, that for any two Functions *i* and *j*, there is a sophisticated relationship governing the occurrences of the start of Intervals along the t-Axis, where during such Intervals, both Functions are simultaneously *high* for duration greater than $x - \mu_1$, where $\mu_1 > 0$ and $(\mu_1/x) \to 0$. This sophisticated relationship involves $T_i$, $T_j$, $D_i$, $D_j$, $\mu_1$, and $x$. Let this relationship be expressed as $G_{ij} = 0$. Next, as there are infinite such Intervals, whose starting points have no constant period, the degree of $G_{ij}$ is also infinite. Therefore when we start considering some another Function *k* defined in Section 2.1 (apart from already considering Functions *i* and *j*), then it is impossible to express Function *k* as a rational multiple of $G_{ij} = 0$. Thus, the Function $G_{ij}=0$, and Function *k*, should again have an infinite number of Intervals, where $x - \mu_2$, where $\mu_2 > 0$ and $(\mu_2/x) \to 0$. Of course, the Intervals while considering Functions *i*, *j*, and *k*, will be less frequent along the t-Axis, than the Intervals when only the Functions *i* and *j* were considered, but there will still be an infinite number of such Intervals. Similarly, as we go on considering more Functions from Section 2.1, one by one, we should find Infinite such Intervals along the t-Axis, though their occurrence is with lesser and lesser frequency.

Angle-2: Let *n* Trains stand next to each other on parallel tracks, with each Train having an infinite number of carriages. Let the length of all the carriages of Train *i* is set to $T_i$, for all *i* in [*1,n*], where the irrational R is very small (*e.g.* $10^{-n} \sqrt{2}$). Each carriage has exactly one window situated at a distance of $D_i$ from the start of that carriage.

Now imagine a window-cleaner-man on Train *1*, who, after cleaning the window of a carriage, moves to clean the window of the next forward carriage of the same Train *1*. Let the man keep moving forward in this way, cleaning windows of the carriages of Train *1*. It is obvious from Theorems 1 and 2, that every time the man looks out from a different window of Train 1, the man will always perceive a different permutation of arrangements of carriages of the parallel Trains *2* to *n*. This also means that if, for every Train, if we were to divide each carriage into *M* Equal parts (where *M* is an extremely large natural number), and color the parts with *M* different colors (same coloring sequence is followed for the *M* parts of all carriages of all Trains), then the following statement is true: - Irrespective of how large *M* is, it is always possible to draw $M^N$ lines orthogonal to the Trains, where each line has a different permutation of colors.

Thus, the fact that **the periods of all *n* Periodic Functions of Section 2.1, are irrational with respect to each other** is the most important factor, which enables us conclude that there exist infinite Intervals along the t-Axis, where all *n* Functions are simultaneously high for a duration greater than $x - \mu$, where $\mu > 0$ and $(\mu/x) \to 0$. **Hence Proved Theorem-4.**

## 3. Our New Approach

### 3.1 Basic Idea of our Model
The basic idea of our Model is to build a set of Equations describing the HCP, such that this set of Equations has a constant number of variables, irrespective of $N$, the number of nodes in the HCP. The feasible solution of each variable is bounded within the space of 2 real numbers.

### 3.2 A Crucial substitution, based on Theorem-4, to keep a Constant number of variables in our Model
For each of the N(N-1) edge-variables in the directed Graph G, we make the following substitution for $x_{i,j}$ (edge-variable for the edge from node $i$ to node $j$, which in our Model, is allowed to have a value that is either approx 2 or approx 0):

$$\forall \{i, j\} \text{ in } [1, N], \text{ and } i \neq j, \text{ we have } x_{i,j} = (1 + \cos(\alpha + ((i-1)N + j)\beta))C_{i,j}$$

$C_{i,j}$ is the cost from Node $i$ and Node $j$, which can be either 0 or 1, and is obtained from the specific instance of the HCP. The above substitution of $x_{i,j}$ is performed with the trigonometric cosine function with period = $1/(1+((i-1)N + j)(\beta/\alpha))$, because from Theorem-4 (replacing $R$ with $\beta/\alpha$), $2^{N(N-1)}$ permutations are possible, as explained by the following statement:

The highest-point/lowest-point of $x_{1,1}$ can come 'extremely close' to the        (statement starts and is contd in next line)
the highest-point/lowest-point of $x_{1,2}$, which can come 'extremely close' to the        (statement contd in next line)
……………….        (statement contd in next line)
the highest-point/lowest-point of $x_{i,j}$, which can come 'extremely close' to the        (statement contd in next line)
……………….        (statement contd in next line)
the highest-point/lowest-point of $x_{n,(n-1)}$, which can come 'extremely close' to the        (statement contd in next line)
the highest-point/lowest-point of $x_{n,n}$.        (statement ends)

By 'extremely close', we mean 'within a distance that tends to 0, and which still may not yet be equal to 0'.

$$Now, \forall\{i, j\} \text{ in } [1, N], \text{ we shall try to express } x_{i,j} \text{ as a polynomial in } \cos(\alpha), \sin(\alpha), \cos(\beta) \text{ and } \sin(\beta).$$

We know from standard Trigonometry that the following 3 Equations are true:
$$\cos(\alpha + k\beta) = \cos(\alpha)\cos(k\beta) - \sin(\alpha)\sin(k\beta), \text{ and}$$
$$\cos(k\beta) = 2\cos((k-1)\beta)\cos(\beta) - \cos(k-2)\beta, \text{ and}$$
$$\sin(k\beta) = 2\sin((k-1)\beta)\cos(\beta) - \sin(k-2)\beta$$

Iteratively follow the above 3 Equations, so as to express $\cos(k\beta) \text{ and } \sin(k\beta)$, as polynomials in $\cos(\beta) \text{ and } \sin(\beta)$. Note that this has been discussed in further detail in [1][2]. It is important, and worthwhile to note that though the co-efficients of the above polynomials have values of the form $2^{P(N)}$ (where $P(N)$ is some polynomial function of $N$), still these co-efficients occupy polynomial size proportional to $log (2^{P(N)})$, which is $P(N)$.

$$Thus, \forall\{i, j\} \text{ in } [1, N], \text{ we are able to express } x_{i,j} \text{ as polynomials in } \cos(\alpha), \sin(\alpha), \cos(\beta) \text{ and } \sin(\beta).$$

The purpose of performing this substitution is to ensure that evaluating the expressions of our Model are maintained within polynomial complexity. For example, it is obvious that, without our substitution, the process of bringing the expression

$$\prod_{i=1}^{N} \left( \sum_{j=1, j \neq i}^{N} (x_{i,j}) \right)$$

into simplest form (i.e. where all terms of the expression have only '+', '-', 'x' or '/' operators) needs exponential space and time. Whereas, our crucial substitution enables this expression to be brought into simplest form within polynomial space and time.

### 3.3 Set of Constraints for our Algorithm (note that before bringing any of these Constraint Sets to Simplest form, we should first Substitute all $x_{i,j}$ by the trigonometric polynomials in 4 variables, as described in Section 3.2)

$$\sum_{i=1}^{N}\left(\sum_{j=1, j\neq i}^{N}\left((2-x_{i,j})^2 + (x_{i,j}-0)^2\right)\right) = 4N(N-1) - \epsilon_1$$

**CONSTRAINT-SET-1**

$$\sum_{i=1}^{N}\left(\left(\prod_{j=1, j\neq i}^{N}(x_{i,j}+1)\right) - 3\right)^2 = \epsilon_2$$

**CONSTRAINT-SET-2**

$$(Y_{1,N+1} - 2)^2 + \sum_{i=2}^{N}(Y_{i,N+1} - 1)^2 = \epsilon_3$$

**CONSTRAINT-SET-3**

Where $Y_{1,1} = 1$, $Y_{i,1} = 0 \ \forall \ i \in [2, N]$, and

$\forall \ i \ in \ [1, N], \ we \ have \ Y_{i,t+1} = Y_{i,t} + \sum_{j=1, j\neq i}^{N}\left((Y_{j,t})(x_{i,j})/2\right)$

$0 \leq \epsilon_1 \leq (2^{-N^2})$
$0 \leq \epsilon_2 \leq (2^{-N^2})$
$0 \leq \epsilon_3 \leq (2^{-N^2})$

**CONSTRAINT-SET-4**

$-1 \leq \cos(\alpha) \leq 1$
$-1 \leq \sin(\alpha) \leq 1$
$-1 \leq \cos(\beta) \leq 1$
$-1 \leq \cos(\beta) \leq 1$
$\sin^2(\alpha) + \cos^2(\alpha) = 1$
$\sin^2(\beta) + \cos^2(\beta) = 1$

**CONSTRAINT-SET-5**

### 3.4 Explanation of the Constraint Sets

In sub-Section 3.3, Constraint Set 5 ensures that each edge variable is within [0,2], and also validates the trigonometric substitution. Constraint Set 1 ensures that the value of each edge-variable is either approximately 2 or approximately 0 (by the word 'approximately' we mean plus/minus a small quantity much less than $2^{-(N \times N)}$). Constraint Set 2 then ensures that there is exactly one edge-variable connected to node *i*, whose value is approximately 2, while all other edge-variables connected to node *i* have values approximately 0. Once this is done, the resulting feasible solutions are either a Hamiltonian Cycle (i.e. Cycle of length = *N*), or isolated rings (i.e. isolated Cycles each of length < (N-2) ).

Next, Constraint Set 3 constrains the previously discussed solutions to only Hamiltonian Cycles (here we assume *N*>2). Constraint Set 3 is actually the simulation of a signal propagating among all the Nodes, from time or t = 1 to t = N, where t increases discretely in steps of 1. At time =1, we start with Unit signal strength at Node 1, and zero signal strength at all the other Nodes. At the next Unit of time, each Node receives signals from all other Nodes, along the edges, and adds it to whatever signal strength it already had. Finally at t = (N+1), if Node 1 has signal strength of approximately 2, and every other Node has a signal strength of approximately 1, then it signifies a Hamiltonian Cycle.

Lastly, Constraint Set 4 constrains all approximation to be within $2^{-(N \times N)}$, so that Constraint Sets 1-3 remain valid. Note that we could have chosen any other number such as $2^{-(N \times N \times N \times N)}$ to constrain the approximation. The reason why we stuck with $2^{-(N \times N)}$, is because it is sufficiently small to ensure that the Model is valid, and still can be expressed within polynomial space of (NxN) bits. Note that a too small number like $2^{-(N \times N \times N \ldots N \text{ times})}$ will take up exponential space.

In summary, **our Constraint Equations are basically looking for the possibility that *N* edge-variables possess values approximately equal to and less than 2, whose corresponding edges form a Hamiltonian Cycle, while all other edge-variables possess values approximately equal to and greater than 0**.

### 3.5 Converting Inequations into Equations

The Inequations described under Section 3.3 (Constraint Sets 4 and 5) can be converted into Equations, because an Inequation of the form $x \leq \lambda$ can be converted into an Equation (by the introduction of an extra variable *z*) of the form: $z^2 + (x - \lambda) = 0$, and then checking for the existence of solutions for both variables *x* and *z*. It is to be noted, that even the extra variable will also be bounded within the space of 2 real numbers.

### 3.6 Final Set of Equations, and their conversion into a single Polynomial Equation

After converting the Inequations to Equations, we finally have a constant number of Polynomial Equations, with a constant number of variables. It is easy to convert them into a single Polynomial Equation, by squaring and adding the Polynomial portions of the Equations, and then equating the result to zero.

## 4. Conclusion, and Future Work

In this paper, we presented a polynomial Approach for expressing the decision version of the directed Hamiltonian Cycle Problem (NP-Complete) or HCP, in terms of the Solvability of an Equation. Using a crucial trigonometric substitution that enables the *N(N-1)* edge-variables to be expressed as polynomials in 4 variables, our Approach describes the HCP with a single Polynomial Equation with a constant number of variables, where the feasible solution of each variable is bounded within a set of 2 real numbers. It is hoped that our approach will be of use, in mankind's P? =NP Question.

We hope someone in the world can develop **EITHER**:

**Technique-1:** A Technique that determines whether or not a Polynomial Equation (with a constant number of variables) is solvable within the Infinite unbounded Real Set, **OR**

**Technique-2:** A Technique that determines whether or not a Polynomial Equation (with a constant number of variables) is solvable within a bounded Real Set (i.e. where the feasible set of each variable is bounded within a space of 2 real numbers)

We could use either of these Techniques for our HCP's Polynomial Equation (Technique-1 will also work for our Model, because we have explicit constraints limiting our variables within the feasible bounded Real Set). So we could use either of the Techniques to determine whether, for a particular HCP instance, if there does exist atleast one solution point for our variables (i.e. *sin(α), cos(α), sin(β), cos(β)*, $\varepsilon_1$, $\varepsilon_2$, $\varepsilon_3$, and the other variables introduced by converting Inequations to Equations), then we can conclude that the HCP instance does have a Hamiltonian Cycle. And if their Technique finds that no such solution is possible, then we can conclude that the HCP instance does not have a Hamiltonian Cycle.

So mankind will be able to prove that P=NP, if either Technique-1 or Technique-2 is developed.

**Acknowledgement**
I am grateful to Dr. David Moews (djm.cc/dmoews.html), who is an IMO Gold Medallist, and three time Putnam Prize Winner with a perfect score in the 1987 difficult one, for his enlightening discussions, and also for pointing out a crucial error in a previous version of my paper. I asked Dr. Moews whether I could co-author him in this paper, but he has not responded yet.